\documentclass[twocolumn]{aastex61}
\pdfoutput=1 %for arXiv submission
\usepackage{amsmath,amstext}
\usepackage[T1]{fontenc}
\usepackage{apjfonts}
\usepackage{comment}
\usepackage{hyperref}
\usepackage{url}
\usepackage{bm}

\usepackage{booktabs}

\shorttitle{A New Limit on CMB Circular Polarization from SPIDER}
\shortauthors{SPIDER Collaboration}

\begin{document}

\title{A New Limit on CMB Circular Polarization from SPIDER}

\newcommand\CWRU{Physics Department, Case Western Reserve University, 10900 Euclid Ave, Rockefeller Building, Cleveland, OH 44106, USA}
\newcommand\Cardiff{School of Physics and Astronomy, Cardiff University, The Parade, Cardiff, CF24 3AA, UK}
\newcommand\UBC{Department of Physics and Astronomy, University of British Columbia, 6224 Agricultural Road,
Vancouver, BC V6T 1Z1, Canada}
\newcommand\Princeton{Department of Physics, Princeton University, Jadwin Hall, Princeton, NJ 08544, USA}
\newcommand\Caltech{Division of Physics, Mathematics and Astronomy, California Institute of Technology, MS 367-17, 1200 E. California Blvd., Pasadena, CA 91125, USA}
\newcommand\JPL{Jet Propulsion Laboratory, Pasadena, CA 91109, USA}
\newcommand\CITA{Canadian Institute for Theoretical Astrophysics, University of Toronto, 60 St. George Street, Toronto, ON M5S 3H8, Canada}
\newcommand\ASU{School of Earth and Space Exploration, Arizona State University, 781 S Terrace Road, Tempe, AZ 85287, USA}
\newcommand\UKZN{School of Mathematics, Statistics and Computer Science, University of KwaZulu-Natal, Durban, South Africa}
\newcommand\NITP{National Institute for Theoretical Physics (NITheP), KwaZulu-Natal, South Africa}
\newcommand\Imperial{Blackett Laboratory, Imperial College London, SW7 2AZ, London, UK}
\newcommand\Stockholm{The Oskar Klein Centre for Cosmoparticle Physics, Department of Physics, Stockholm University, AlbaNova, SE-106 91 Stockholm, Sweden}
\newcommand\Oslo{Institute of Theoretical Astrophysics, University of Oslo, P.O. Box 1029 Blindern, NO-0315 Oslo, Norway}
\newcommand\Toronto{Department of Astronomy and Astrophysics, University of Toronto, 50 St George Street, Toronto, ON M5S 3H4 Canada}
\newcommand\UIUCP{Department of Physics, University of Illinois at Urbana-Champaign, 1110 W. Green Street, Urbana, IL 61801, USA}
\newcommand\UIUCA{Department of Astronomy, University of Illinois at Urbana-Champaign, 1002 W. Green Street, Urbana, IL 61801, USA}
\newcommand\NRAO{National Radio Astronomy Observatory, Charlottesville, NC 22903, USA}
\newcommand\Michigan{Department of Physics, University of Michigan, 450 Church Street, Ann  Arbor, MI 48109, USA}
\newcommand\TorontoP{Department of Physics, University of Toronto, 60 St George Street, Toronto, ON M5S 3H4 Canada}
\newcommand\Hopkins{Department of Physics and Astronomy, Johns Hopkins University, 3400 N. Charles Street, Baltimore, MD 21218 USA}
\newcommand\Goddard{NASA Goddard Space Flight Center, Code 665, Greenbelt, MD 20771, USA}
\newcommand\APC{APC, Univ. Paris Diderot, CNRS/IN2P3, CEA/Irfu, Obs de Paris, Sorbonne Paris Cit\'e, France}
\newcommand\PennState{Department of Astronomy and Astrophysics, Pennsylvania State University, 520 Davey Lab, University Park, PA 16802, USA}
\newcommand\NIST{National Institute of Standards and Technology, 325 Broadway Mailcode 817.03, Boulder, CO 80305, USA}
\newcommand\Stanford{Department of Physics, Stanford University, 382 Via Pueblo Mall, Stanford, CA 94305, USA}
\newcommand\SLAC{ SLAC National Accelerator Laboratory, 2575 Sand Hill Road, Menlo Park, CA 94025, USA}
\newcommand\PrincetonEngineering{Department of Mechanical and Aerospace Engineering, Princeton University, Engineering Quadrangle, Princeton, NJ 08544, USA}
\newcommand\Fermilab{Fermi National Accelerator Laboratory, P.O. Box 500, Batavia, IL 60510-5011, USA}
\newcommand\KICPChicago{Kavli Institute for Cosmological Physics, University of Chicago, 5640 S Ellis Avenue, Chicago, IL 60637 USA}
\newcommand\Orsay{Institut d'Astrophysique Spatiale, Orsay, France}
\newcommand\MPI{Max-Planck-Institute for Astronomy, Konigstuhl 17, 69117, Heidelberg, Germany}
\newcommand\LAIM{Laboratoire AIM, Paris-Saclay, CEA/IRFU/SAp - CNRS - Universit\'e Paris Diderot, 91191, Gif-sur-Yvette Cedex, France}

\author{ J.~M.~Nagy }
\affiliation{\CWRU}
\email{johanna.nagy@case.edu}

\author{ P.~A.~R.~Ade }
\affiliation{\Cardiff}

\author{ M.~Amiri }
\affiliation{\UBC}

\author{ S.~J.~Benton }
\affiliation{\Princeton}

\author{ A.~S.~Bergman }
\affiliation{\Princeton}

\author{ R.~Bihary }
\affiliation{\CWRU}

\author{ J.~J.~Bock }
\affiliation{\Caltech}
\affiliation{\JPL}

\author{ J.~R.~Bond }
\affiliation{\CITA}

\author{ S.~A.~Bryan }
\affiliation{\ASU}

\author{ H.~C.~Chiang }
\affiliation{\UKZN}
\affiliation{\NITP}

\author{ C.~R.~Contaldi }
\affiliation{\Imperial}

\author{ O.~Dor{\'e} }
\affiliation{\Caltech}
\affiliation{\JPL}

\author{ A.~J.~Duivenvoorden }
\affiliation{\Stockholm}

\author{ H.~K.~Eriksen }
\affiliation{\Oslo}

\author{ M.~Farhang }
\affiliation{\CITA}
\affiliation{\Toronto}

\author{ J.~P.~Filippini }
\affiliation{\UIUCP}
\affiliation{\UIUCA}

\author{ L.~M.~Fissel }
\affiliation{\NRAO}
\affiliation{\Toronto}

\author{ A.~A.~Fraisse }
\affiliation{\Princeton}

\author{ K.~Freese }
\affiliation{\Michigan}
\affiliation{\Stockholm}

\author{ M.~Galloway }
\affiliation{\TorontoP}

\author{ A.~E.~Gambrel }
\affiliation{\Princeton}

\author{ N.~N.~Gandilo }
\affiliation{\Hopkins}
\affiliation{\Goddard}

\author{ K.~Ganga }
\affiliation{\APC}

\author{ J.~E.~Gudmundsson }
\affiliation{\Stockholm}

\author{ M.~Halpern }
\affiliation{\UBC}

\author{ J.~Hartley }
\affiliation{\TorontoP}

\author{ M.~Hasselfield }
\affiliation{\PennState}

\author{ G.~Hilton }
\affiliation{\NIST}

\author{ W.~Holmes }
\affiliation{\JPL}

\author{ V.~V.~Hristov }
\affiliation{\Caltech}

\author{ Z.~Huang }
\affiliation{\CITA}

\author{ K.~D.~Irwin }
\affiliation{\Stanford}
\affiliation{\SLAC}

\author{ W.~C.~Jones }
\affiliation{\Princeton}

\author{ C.~L.~Kuo }
\affiliation{\Stanford}

\author{ Z.~D.~Kermish }
\affiliation{\Princeton}

\author{ S.~Li }
\affiliation{\Toronto}
\affiliation{\Princeton}
\affiliation{\PrincetonEngineering}

\author{ P.~V.~Mason }
\affiliation{\Caltech}

\author{ K.~Megerian }
\affiliation{\JPL}

\author{ L.~Moncelsi }
\affiliation{\Caltech}

\author{ T.~A.~Morford }
\affiliation{\Caltech}

\author{ C.~B.~Netterfield }
\affiliation{\Toronto}
\affiliation{\TorontoP}

\author{ M.~Nolta }
\affiliation{\CITA}

\author{ I.~L.~Padilla }
\affiliation{\Toronto}

\author{ B.~Racine }
\affiliation{\Oslo}

\author{ A.~S.~Rahlin }
\affiliation{\Fermilab}
\affiliation{\KICPChicago}

\author{ C.~Reintsema }
\affiliation{\NIST}

\author{ J.~E.~Ruhl }
\affiliation{\CWRU}

\author{ M.~C.~Runyan }
\affiliation{\JPL}

\author{ T.~M.~Ruud }
\affiliation{\Oslo}

\author{ J.~A.~Shariff }
\affiliation{\CITA}

\author{ J.~D.~Soler }
\affiliation{\MPI}
\affiliation{\LAIM}

\author{ X.~Song }
\affiliation{\Princeton}

\author{ A.~Trangsrud }
\affiliation{\Caltech}
\affiliation{\JPL}

\author{ C.~Tucker }
\affiliation{\Cardiff}

\author{ R.~S.~Tucker }
\affiliation{\Caltech}

\author{ A.~D.~Turner }
\affiliation{\JPL}

\author{ J.~F.~van~der~List }
\affiliation{\Princeton}

\author{ A.~C.~Weber }
\affiliation{\JPL}

\author{ I.~K.~Wehus }
\affiliation{\Oslo}

\author{ D.~V.~Wiebe }
\affiliation{\UBC}

\author{ E.~Y.~Young }
\affiliation{\Princeton}

% \clearpage

\begin{abstract}
We present a new upper limit on CMB circular polarization from the 2015 flight of \textsc{Spider}, a balloon-borne telescope designed to search for $B$-mode linear polarization from cosmic inflation. Although the level of circular polarization in the CMB is predicted to be very small, experimental limits provide a valuable test of the underlying models. By exploiting the non-zero circular-to-linear polarization coupling of the HWP polarization modulators, data from \textsc{Spider}'s 2015 Antarctic flight provide a constraint on Stokes $V$ at 95 and 150 GHz from $33<\ell<307$. No other limits exist over this full range of angular scales, and \textsc{Spider} improves upon the previous limit by several orders of magnitude, providing 95\% C.L. constraints on $\ell (\ell+1)C_{\ell}^{VV}/(2\pi)$ ranging from 141 $\mu K ^2$ to 255 $\mu K ^2$ at 150 GHz for a thermal CMB spectrum. As linear CMB polarization experiments become increasingly sensitive, the techniques described in this paper can be applied to obtain even stronger constraints on circular polarization.
\end{abstract}
%\keywords{keyword1 --- keyword2 --- keyword3}

\section{Introduction}
\label{sec:intro}
Anisotropies in the intensity and linear polarization of the Cosmic Microwave Background (CMB) have provided a wealth of information about the history and contents of the universe.  Standard cosmological models do not predict a measurable amount of circular polarization, characterized by the Stokes $V$ parameter, in the CMB;  as such, any detection of a primordial $V$ signal would be of enormous interest.   A variety of secondary physical processes may produce circular polarization in the CMB at very low levels. For instance, Faraday conversion can transform existing linear polarization into circular polarization in both the magnetic fields of galaxy clusters \citep{galaxy_faraday} and the relativistic plasma remnants of Population III stars \citep{pop3_stars}.  Magnetic fields in the primordial universe \citep{giovannini_primordial, zarei_primordial}, scattering from the cosmic neutrino background \citep{CvB}, and photon-photon interactions in neutral hydrogen \citep{photon_photon} have also been shown to potentially produce CMB circular polarization. Additional sources include postulated extensions to QED such as Lorentz-invariance violating operators \citep{QED2a, QED2b}, axion-like pseudoscalar particles \citep{axions}, and non-linear photon interactions (through effective Euler-Heisenberg Lagrangians) \citep{QED1}.  A brief review of some of these generation mechanisms can be found in \citet{Lubin_review}.  Despite the wide range of physical processes they invoke, all of these mechanisms predict levels of circular polarization that are unlikely to be accessible with current technology.

Nevertheless, circular polarization measurements provide a valuable test of the standard cosmological model and the physics behind these generation mechanisms.  Yet there are relatively few published limits.  MIPOL reported the strongest constraint on large angular scales ($\ell < 30$) in 2013, providing  95\% C.L. measurements ranging from $\Delta V/T_{\mathrm{CMB}}\leq 2.4 \times 10^{-4} $ to $\Delta V/T_{\mathrm{CMB}}\leq 4.3 \times 10^{-4} $  at 33 GHz \citep{Mipol}. This is roughly an order of magnitude better than the previous 95\% C.L. limit of $\Delta V/T_{\mathrm{CMB}}\leq 4 \times 10^{-3} $ at 33 GHz at $\ell \approx 10$ \citep{Vlim2}.  On smaller angular scales, the only reported measurement comes from the VLA, which set 95\% C.L. limits at 5 GHz between $\Delta V/T_{\mathrm{CMB}}\leq 2.2 \times 10^{-4} $ and $\Delta V/T_{\mathrm{CMB}}\leq 0.6 \times 10^{-4} $ for a range of angular scales with $\ell > 3000$ \citep{Vlim3}. 

These limits are more than 7 orders of magnitude higher than the best measurements of the linear polarization power spectra, but there are no contemporary experiments with the primary goal of improving them.  However, some modern linear polarization experiments, such as \textsc{Spider}, can take advantage of this vast disparity to set stronger limits as a consequence of their polarization modulation techniques. 

\textsc{Spider} is a balloon-borne CMB telescope that is searching for a $B$-mode linear polarization signal from cosmic inflation \citep{Fraisse, Rahlin2014}.  During its first flight in January 2015, \textsc{Spider} made maps of approximately 10\% of the sky with degree-scale angular resolution in 95 and 150 GHz observing bands.  The analysis of the linear polarization data from this flight is currently in progress.  In this paper, we exploit non-idealities of \textsc{Spider}'s half-wave plate (HWP) polarization modulators to obtain a new upper limit on CMB circular polarization. 

The \textsc{Spider} payload features six monochromatic receivers housed in a shared cryostat \citep{cryo_paper}.  Each receiver includes a stepped HWP polarization modulator to reduce the potential impact of systematic errors due to beam asymmetries and instrument polarization \citep{Sean_thesis, mechanism_paper}.  Although \textsc{Spider}'s antenna-coupled TES bolometers are not sensitive to variations in circular polarization \citep{JPL_detectors}, non-idealities of the HWPs allow a measurement of the Stokes $V$ parameter after combining maps made at several HWP angles.  The calculation of \textsc{Spider}'s circular polarization coupling is described in the next section.  Section \ref{sec:limit} details how this coupling is used to derive a circular polarization limit.  The significance of this result and prospects for future measurements are discussed in Section \ref{sec:discussion}.

\section{Coupling to Circular Polarization}
A birefringent material forms a half-wave plate when the difference in the optical path length between waves polarized along the fast and slow crystal axes is exactly half of the photon wavelength.  An ideal HWP rotates the polarization plane of the light passing through it by $2\theta_{HWP}$, where $\theta_{HWP}$ is the angle between the incoming polarization plane and the slow crystal axis.  However, this condition can only be exactly satisfied at a single frequency.  Similarly, the single-layer anti-reflection (AR) coatings applied to \textsc{Spider}'s HWPs are not uniformly efficient over the observing bands.  When combined, these conditions lead to a frequency-dependent reduction in transmission through the HWPs and induce non-ideal polarization modulation effects as the HWPs are rotated.   Since \textsc{Spider}'s observing bands have roughly 20\% bandwidths, the magnitude of such effects could be significant. 

Following \citet{bryan2010modeling}, a non-ideal HWP can be modeled with four parameters that can be broadly interpreted as the total transmission $T$, the difference in transmission between the fast and slow axes $\rho$, the linear polarization response $c$, and the coupling to circular polarization $s$.  In terms of these parameters, the Mueller matrix of a HWP with its birefringent crystal axes oriented along the horizontal and vertical directions can be written as
\begin{equation}\label{m_hwp}\arraycolsep=1.0pt\def\arraystretch{1.0}
M_{HWP} = \left[ \begin{array}{cccc} T & \rho & 0 & 0 \\ \rho & T & 0 & 0  \\ 0 & 0 & c & -s \\ 0 & 0 & s & c \end{array} \right].
\end{equation} 
For an ideal HWP, $T=1$, $c=-1$, and $\rho=s=0$.  The ideal case captures the effect of the HWP on linear polarization signals and has no coupling between linear and circular polarization ($s = 0$).  In real HWPs, however, these parameters can deviate significantly from their ideal values. While simulations have shown that these non-idealities are not problematic for detecting a $B$-mode signal at \textsc{Spider'}s anticipated sensitivity level \citep{Spider_odea}, they allow \textsc{Spider} to measure circular polarization to the extent that $s$ is non-zero.

The sky signal in a detector timestream $d$ is given in terms of the Stokes parameters $I$, $Q$, $U$, and $V$ and the instrument Mueller matrix elements $M_{XY}$ by
\begin{equation}\label{d_eq}
d = IM_{II} + QM_{IQ} + UM_{IU} + VM_{IV}.
\end{equation}
The instrument Mueller matrix is calculated in \citet{bryan2010modeling} by multiplying the Mueller matrices of every element in the optical chain, including $M_{HWP}$ from Equation \ref{m_hwp}.  For the purposes of this paper we are interested only in the instrument Mueller matrix element $M_{IV}$.  Generalizing the treatment in \citet{bryan2010modeling} for arbitrary detector angles, it is straightforward to show that the $V$ parameter couples to a detector timestream as

\begin{equation}\label{MIV}
M_{IV} = s\gamma \sin(2\theta_{HWP} - 2 \xi_{det}).
\end{equation}

Here $\theta_{HWP}$ is the HWP angle and $\xi_{det}$ is the detector angle, both of which are defined relative to the instrument.  Note that $M_{IV}$ does not depend on the rotational orientation of the instrument relative to the sky and can be positive, negative, or zero depending on the relative HWP and detector angles.  The overall polarization efficiency of the instrument is described by $\gamma$, while $s$ describes the coupling to circular polarization from the HWP non-idealities.  Note that $s$ does not appear in the $M_{II}$, $M_{IQ}$, or $M_{IU}$ matrix elements in Equation \ref{d_eq} and therefore is not used in \textsc{Spider}'s linear polarization analysis.

\begin{table*}
\begin{center}
    \begin{tabular}{*5l} 
    \toprule
    95 GHz Receivers & X2 & X4 & X6 \\ \midrule
    Sapphire (mm) & 4.97 $\pm$ 0.01  & 4.94 $\pm$ 0.01 & 4.97 $\pm$ 0.01 \\ 
    Top Quartz Layer (mm) & 0.420 $\pm$ 0.015 & 0.429 $\pm$ 0.015 & 0.427 $\pm$ 0.015 \\ 
    Bottom Quartz Layer (mm)  & 0.419 $\pm$ 0.015 & 0.419 $\pm$ 0.015 & 0.422 $\pm$ 0.015 \\
    Gap (mm) & 0.01 $\pm$ 0.01 & 0.01 $\pm$ 0.01  & 0.01 $\pm$ 0.01 \\
    \bottomrule
    \end{tabular}
    \caption{Thicknesses of the 95 GHz HWP components. \textsc{Spider}'s HWPs are made from single-crystal birefringent sapphires, which are AR coated to maximize transmission \citep{Sean_thesis}. At 95 GHz, the sapphires are AR coated with quartz wafers that are glued at the centers and held by spring clips at the edges.  The measured thicknesses of these materials for each HWP are listed in the table.  Since the adhesive covers only a small fraction of the total surface area, it is ignored in the calculation of $s$.  However, the possibility of a narrow gap between the sapphire and the quartz is taken into account.}
    \label{table: HWP_table_90}
\end{center}
\end{table*}

\begin{table*}
\begin{center}
    \begin{tabular}{ *5l } 
    \toprule
    150 GHz Receivers & X1 & X3 & X5 \\ \midrule
    Sapphire (mm) & 3.21 $\pm$ 0.01  & 3.26 $\pm$ 0.01 & 3.23 $\pm$ 0.01 \\ 
    Top Cirlex Layer (mm) &	0.250 $\pm$ 0.005 & 0.250 $\pm$ 0.005 & 0.250 $\pm$ 0.005 \\ 
    Bottom Cirlex Layer (mm)  & 0.250 $\pm$ 0.005 & 0.250 $\pm$ 0.005 & 0.250 $\pm$ 0.005 \\
    HDPE Bond Layer (mm) & 0.006 $\pm$ 0.001 & 0.006 $\pm$ 0.001  & 0.006 $\pm$ 0.001 \\
    \bottomrule
    \end{tabular}
    \caption{Thicknesses of the 150 GHz HWP components. The 150 GHz HWPs are AR coated with Cirlex, a polyimide film, which is adhered with a melted HDPE bond layer.  The measured thicknesses of these materials for each HWP are listed in the table.  The uniformity in the thickness of the Cirlex sheets can likely be attributed to a common production batch.}
    \label{table: HWP_table_150}
\end{center}
\end{table*}

\begin{table*}
\begin{center}
    \begin{tabular}{ *5l } 
    \toprule
    Material & Refractive Index ($n$) & Reference \\ \midrule
    Sapphire (fast axis) & 3.019 $\pm$ 0.003  &	\citet{bryanSPIE}  \\ 
    Sapphire (slow axis) & 3.336 $\pm$ 0.003 &	\citet{bryanSPIE} \\ 
    Quartz (fused)  & 1.95 $\pm$ 0.01 & \citet{Sean_thesis} \\
    Cirlex & 1.94 $\pm$ 0.01 & \citet{Sean_thesis} \\
    HDPE & 1.56 $\pm$ 0.01 & \citet{LambMaterials} \\
    \bottomrule
    \end{tabular}
    \caption{Refractive indices of the HWP materials.  The listed values assume a temperature of approximately 4 K and observing bands in the range of 50-200 GHz.}
    \label{table: n_table}
\end{center}
\end{table*}

\textsc{Spider}'s six receivers are assigned names consisting of the letter `X' followed by the numbers 1 through 6, where the even numbers refer to 95 GHz receivers and the odd numbers to 150 GHz receivers.  Each receiver has a dedicated HWP and therefore a unique value of the $s$ non-ideality parameter.  It can be calculated as described in \citet{bryan2010modeling} from the thicknesses and refractive indices of the HWP materials, the spectrum of the observed source, and the shape of the observing band.  Similar HWP modeling techniques have been used for the linear polarization properties of sapphire HWPs by \citet{Cardiff_modeling1} and found to be in good agreement with experimental measurements \citep{Cardiff_measurement1, SCUBA2_HWP, Sean_thesis}.  

For the results presented in this paper, uncertainties in the component properties lead to significant uncertainty in $s$ for each HWP, which is quantified with Monte Carlo simulations.  We use the temperature derivative of the CMB blackbody spectrum for the source in the baseline case, as well as the thicknesses and uncertainties of the HWP components listed in Tables \ref{table: HWP_table_90} and \ref{table: HWP_table_150}, and the refractive indices and uncertainties of the materials in Table \ref{table: n_table}.  Since the refractive indices of sapphire should be the same for every HWP, we use the same randomly drawn values of the two indices for all receivers in each iteration of the $s$ calculations.

To take \textsc{Spider}'s observing bands into account, we use Fourier Transform Spectrometer (FTS) measurements made just prior to flight \citep{Jon_thesis}.  However, correctly interpreting these measurements relies on knowing the frequency dependence of the coupling to the Rayleigh-Jeans calibration source.  Although the intensity of the source has a $\nu^2$ frequency dependence, the beam throughput ($A\Omega$) scales as $\nu^{-2}$ in the beam-filling limit.  For \textsc{Spider}, the source is not entirely beam filling, and calculations indicate that this coupling should be approximately $\nu^{-1.5}$.  This leads to larger absolute values of $s$ than in the beam-filling case.  However, due to the relatively large uncertainty in the calculation of this scaling, we adopt a conservative approach in this paper and assume a $\nu^{-2}$ coupling, likely underestimating $|s|$.

\begin{figure*}
\begin{centering}
\includegraphics[scale=0.4]{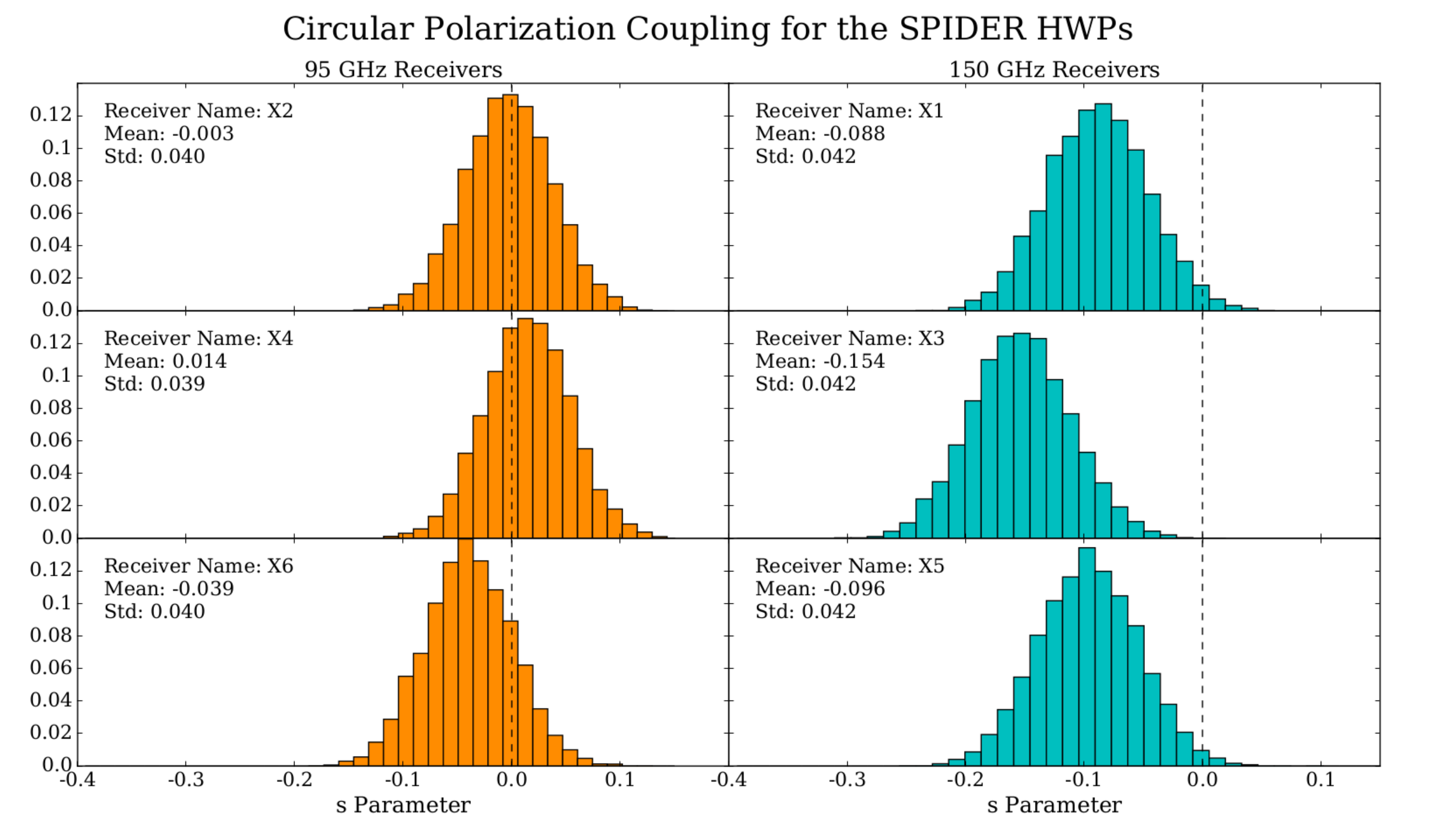}
\par
\end{centering}

\caption{Probability distributions of the $s$ parameter for each \textsc{Spider} receiver.  Each distribution is based on 10,000 Monte Carlo simulations that include a CMB source spectrum, the measured observing band, and the physical properties of the individual HWP.  The 150 GHz HWPs have larger absolute values of $s$ than the 95 GHz HWPs primarily because the sapphire thicknesses are not as well matched to the observing bands.  Although  some  of  the  distributions include $s=0$ with a substantial probability, having three different HWPs at each frequency greatly improves \textsc{Spider}'s statistical power to constrain $V$.}
\label{fig: all_s}
\end{figure*}

The probability distributions of the $s$ parameters for a CMB source for each \textsc{Spider} receiver are shown in Figure \ref{fig: all_s}. These are derived from 10,000 Monte Carlo simulations, which calculate a new value of $s$ for each iteration following the methodology presented by \citet{bryan2010modeling}, using randomly drawn sets of physical parameters based on the central values and uncertainties listed in Tables \ref{table: HWP_table_90}-\ref{table: n_table}.  The distributions of $s$ for the 150 GHz systems exclude zero at roughly 2- to 4-$\sigma$.  At 95 GHz, they include $s=0$ within the 1-$\sigma$ range.  However, to the extent that these distributions are truly good estimators of the $s$ probability distributions, they can still be used to constrain the amplitude of circular polarization by virtue of the significant probability of non-zero $s$.  Note that having three separate HWPs at each frequency greatly improves \textsc{Spider}'s statistical power to constrain $V$.  The differences in distributions between receivers at the same frequency are caused by differences in the shapes of the measured observing bands for each receiver and in the measured thicknesses of the actual HWP components.

\section{Results}
\label{sec:limit}

\begin{figure*}
\begin{centering}
\includegraphics[scale=0.44]{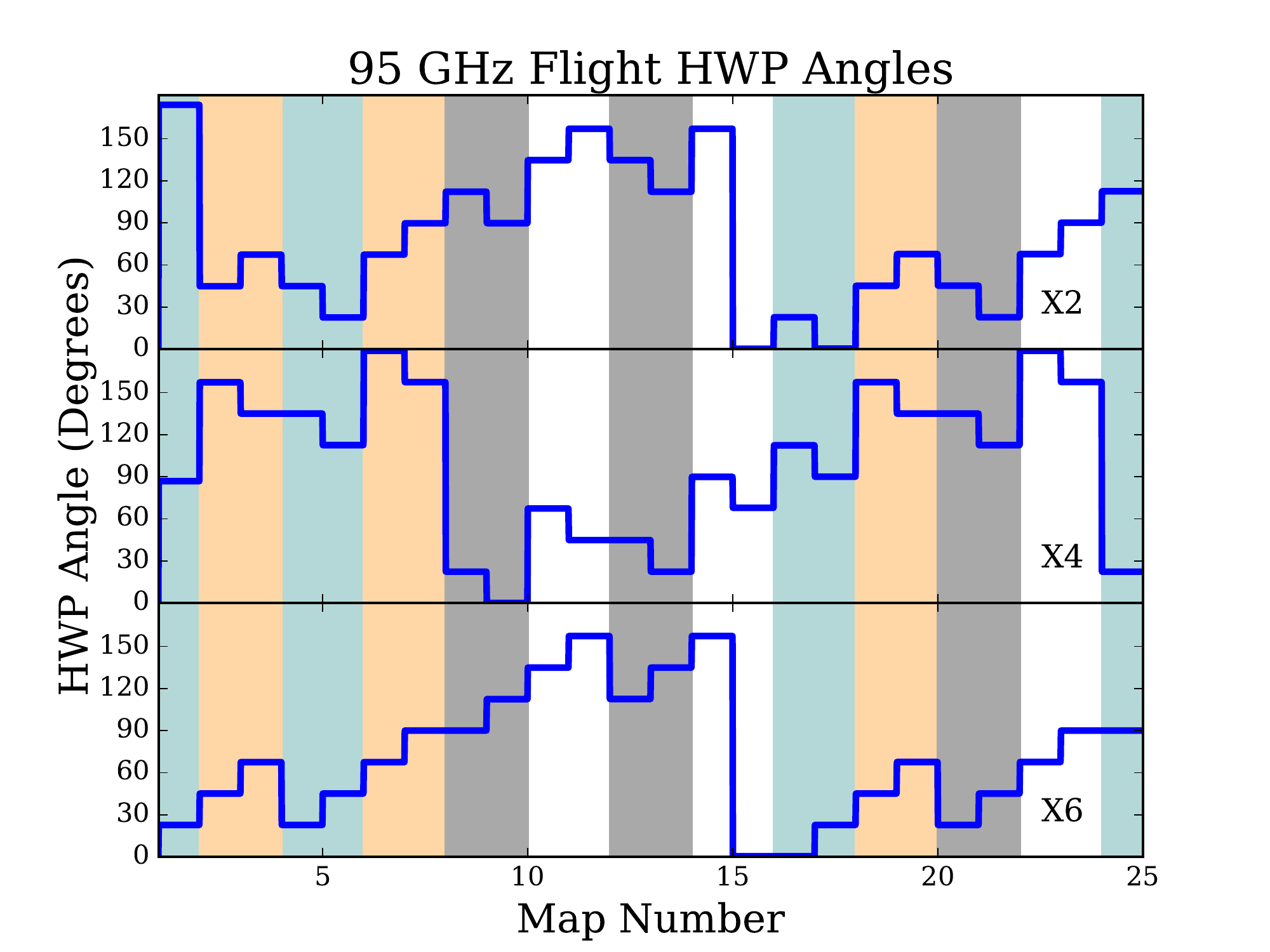}
\includegraphics[scale=0.44]{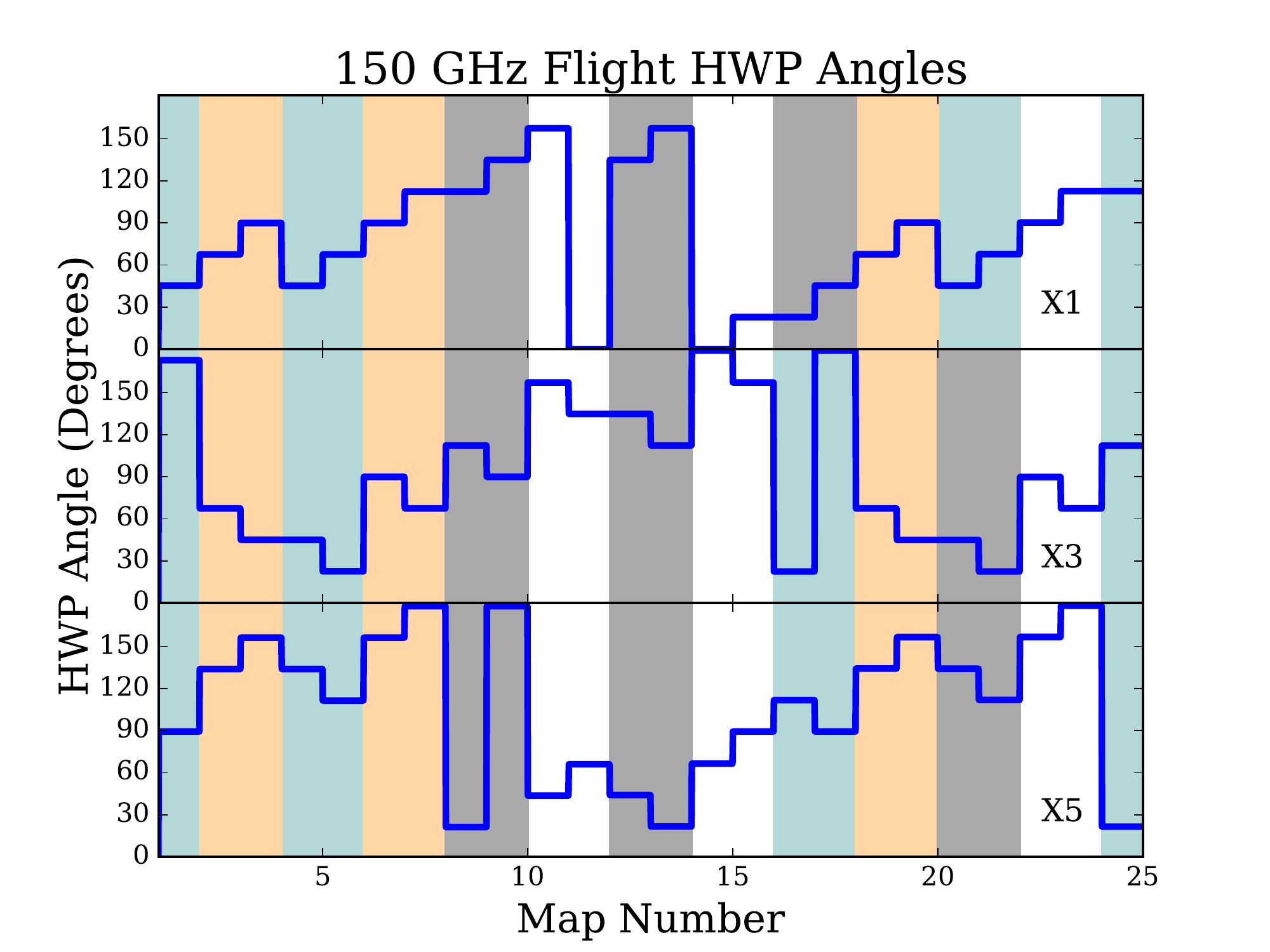}
\par
\end{centering}

\caption{The HWP observing angles from \textsc{Spider}'s 2015 flight.  These angles are defined relative to the slow crystal axis, and the error on the angle difference $2\theta_{HWP} - 2 \xi_{det}$ from Equation~\ref{MIV} is $<1$ degree.  The nominal HWP angles are spaced at integer multiples of 22.5 degrees, and the receivers spent approximately 12 sidereal hours observing at each HWP position, covering the desired region once during that time.  The shaded colors indicate sets of maps on each receiver that were combined to make the cross-spectra described in Section \ref{sec:limit}.  Each set contains an approximately equal amount of data and includes maps made with both wide and narrow scans.  The unique combination of maps on the X1 receiver compensates for an offset of 22.5 degrees from the intended rotation schedule.} 
\label{fig: HWP_angles}
\end{figure*}

During the 2015 flight, \textsc{Spider} observed approximately 4500 square degrees of sky near the southern Galactic pole, centered around roughly RA=50$^\circ$ and Dec=-35$^\circ$. For the first 7.5 days used in this analysis, almost the entire region was mapped every 12 sidereal hours following a sinusoidal azimuth scan profile and using a scan width of $\sim$70$^\circ$ peak-to-peak.  Maps for the remaining 4.5 days covered smaller overlapping sub-regions using narrower sinusoidal azimuth scans with widths of $\sim$35$^\circ$ peak-to-peak \citep{Jamil_thesis}.  The HWPs were held at fixed angles during each of these maps and rotated to new angles between them, following the patterns shown in Figure \ref{fig: HWP_angles}.  Over the course of the flight, each receiver observed the sky at 8 discrete HWP angles nominally spaced at 22.5 degree intervals.

The data from individual receivers are combined into 4 independent sets, illustrated by the colored bands in Figure \ref{fig: HWP_angles}, which were optimized for separating the $Q$, $U$, and $V$ signals.  Each of these sets is used to construct an independent $V$ map with a binned map-maker \citep{Sasha_thesis}, using the values of polarization efficiency $\gamma$ listed in Table \ref{table: gamma}.  If the $s$ value for each receiver was known exactly, Equations \ref{d_eq} and \ref{MIV} could be used by the mapmaker to construct $V$ maps directly from the \textsc{Spider} data.  Instead, since the values of $s$ actually follow broad probability distributions, and $s$ appears only in the $M_{IV}$ matrix element in Equation \ref{d_eq}, we make $V$ maps assuming $s=1$ and later scale the resulting power spectra.

\begin{table}[ht]
\begin{center}
    \begin{tabular}{ *5l } 
    \toprule
    Receiver Name & Frequency & Polarization Efficiency ($\gamma$) \\ \midrule
    X1 & 150 GHz  &	0.959 $\pm$ 0.005 \\ 
    X2 & 95 GHz  &	0.965 $\pm$ 0.001 \\ 
    X3 & 150 GHz  &	0.950 $\pm$ 0.008 \\
    X4 & 95 GHz  &	0.964 $\pm$ 0.001 \\
    X5 & 150 GHz  &	0.956 $\pm$ 0.005 \\
    X6 & 95 GHz  &	0.964 $\pm$ 0.003 \\
    \bottomrule
    \end{tabular}
    \caption{\textsc{Spider}'s polarization efficiency $\gamma$. These values were obtained by combining calculations of the four HWP non-ideality parameters with measurements of the detector cross-polarization response. Since $\gamma$ is dominated by the contribution from the HWPs, the same value is used for every detector on a given receiver.}
    \label{table: gamma}
\end{center}
\end{table}

Before making these maps, glitches such as cosmic ray hits, payload transmitter signals, and thermal transients are identified and removed from the detector timestreams.  This pipeline is shared with the linear polarization analysis and will be described more extensively in a future publication. Some detectors have been excluded from this analysis due to undesirable remaining timestream features, but a number of them may be recovered for future results.  Here we use 681 detectors at 95 GHz and 1117 detectors at 150 GHz, rejecting an average of approximately 30\% of the data from these timestreams.  For this result we subtract a fifth-order polynomial fit in azimuth from each scan (approximately 30 seconds of data) to remove scan-synchronous noise.

Only part of \textsc{Spider}'s observing region is used for the circular polarization analysis, masking data outside of $30^{\circ} \leq \mathrm{RA} \leq 70^{\circ}$ and $-55^{\circ} \leq \mathrm{Dec} \leq -15^{\circ}$.  The leakage from other signals to $V$ is subtracted in map-space from full timestream signal simulations based on Planck 100 and 143 GHz temperature-only input maps \citep{Planck_ref}.  This is dominated by $T$-to-$V$ leakage from the polynomial timestream filter, which is at the level of $\sim 10 \; \mu K$ in the original $s=1$ maps and roughly 2 orders of magnitude smaller than our $V$ sensitivity.  The $E$-to-$V$ leakage is about 5 orders of magnitude lower.  The $V$ pipeline was verified through simulations in which an input signal-only $V$ map was observed following \textsc{Spider}'s scan strategy and then recovered after applying the same flagging and filtering to the re-observed timestreams.

The cross-spectra of the $s=1$ maps are estimated with PolSpice \citep{PolSpice}, which takes the sky mask into account.  We apply a transfer function to account for the effects of timestream filtering and beam smoothing, where the beam correction is derived from map-domain fits to Planck temperature maps.  The transfer function is obtained by comparing $TT$ spectra from smoothed Planck maps of the \textsc{Spider} region to spectra from simulated re-observations of those Planck maps that include \textsc{Spider}'s pointing and timestream filtering.  

The $s=1$ cross-spectra for pairs of maps at a given frequency are then combined with Monte Carlo simulations.  In each iteration, values of $s$ for receivers $i$ and $j$ are drawn from the distributions shown in Figure \ref{fig: all_s}, and the cross-spectra are then scaled by $1/(s_i s_j)$.  Note that the selected $s$ values are slightly correlated due to the common sapphire indices.  We calculate the weighted mean of the resulting values in each $\ell$ bin, weighting by the variance of the $s=1$ map cross-spectra in that bin.  This process is repeated 10,000 times, and the mean and error in each $\ell$ bin are derived from the resulting distribution.

At 150 GHz we cross every pair of maps from each of the three receivers and four independent sets, excluding the auto-spectra, for a total of 66 cross-spectra.  This includes crossing maps made simultaneously on different receivers because the noise has been shown to be no more correlated than in any of the other map pairs.  At 95 GHz we only cross maps from the same receiver because the $s$ distributions allow both positive and negative $s$ values with significant probabilities.  A sign error on either $s_i$ or $s_j$ (but not both) relative to the true value would flip the sign of the cross-spectrum, potentially suppressing real $V$ signals upon averaging.  By restricting ourselves to the 18 cross-spectra that can be constructed from single-receiver maps, we ensure that $s_{i}s_{j}=s_{i}^2$ is always positive, thus avoiding this problem at the price of a small noise penalty.

\begin{figure*}
\begin{centering}
\includegraphics[scale=0.44]{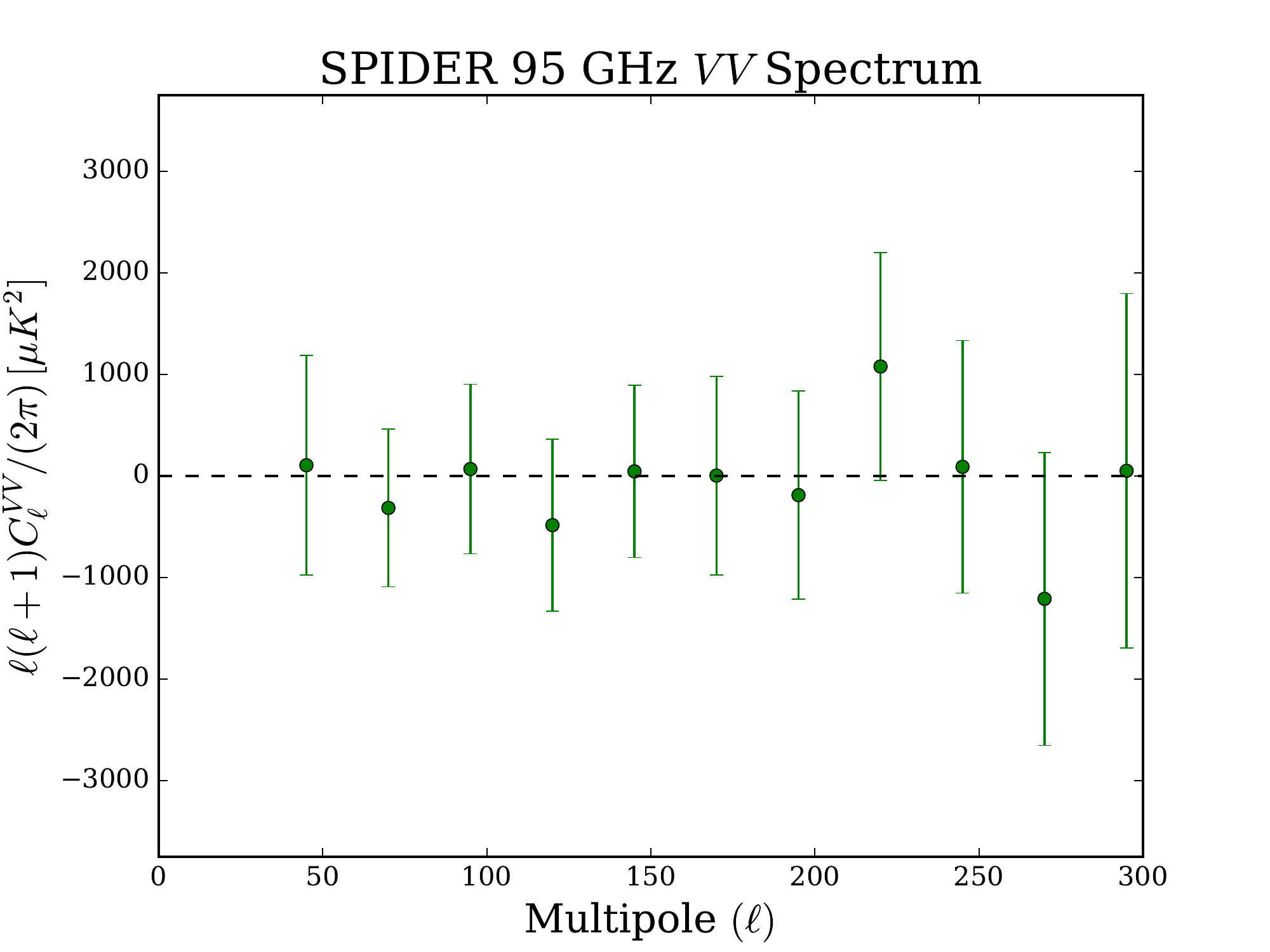}
\includegraphics[scale=0.44]{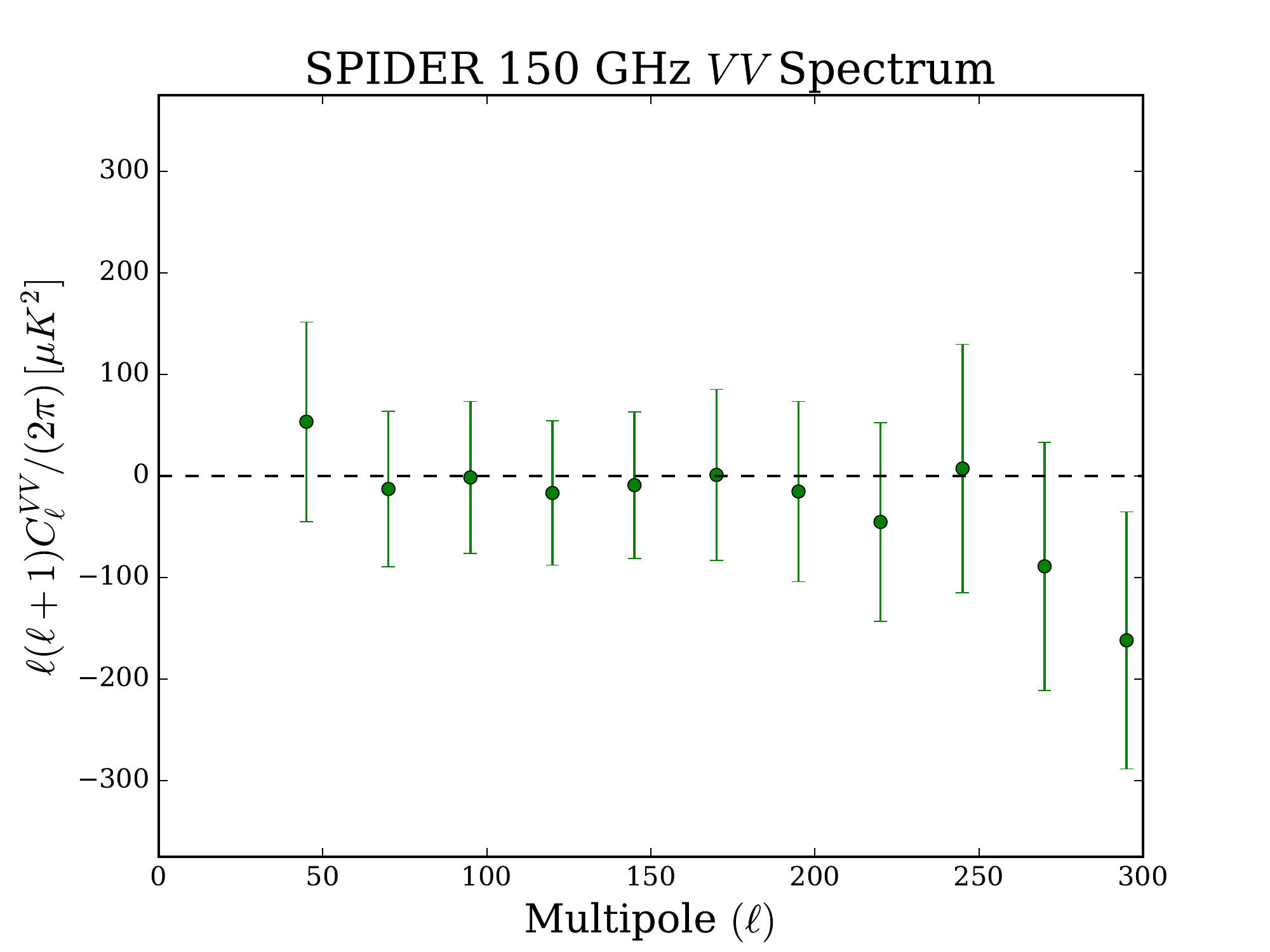}
\par
\end{centering}

\caption{\textsc{Spider}'s $VV$ angular power spectra at 95 and 150 GHz.  The spectra are made by combining data on all three receivers at each frequency and include 68\% C.L. error bars.  The errors are obtained from Monte Carlo simulations based on the spread in the cross-spectra and the uncertainty in the circular polarization coupling of each HWP from Figure \ref{fig: all_s}.  The latter contribution is highly correlated across all $\ell$ bins, leading to the visually low scatter in the points relative to the plotted errors.  Note that the $y$-axis is a factor of 10 larger in the 95 GHz spectrum than in the 150 GHz spectrum.} 
\label{fig: V_spectra}
\end{figure*}

\begin{figure*}
\begin{centering}
\includegraphics[scale=0.75]{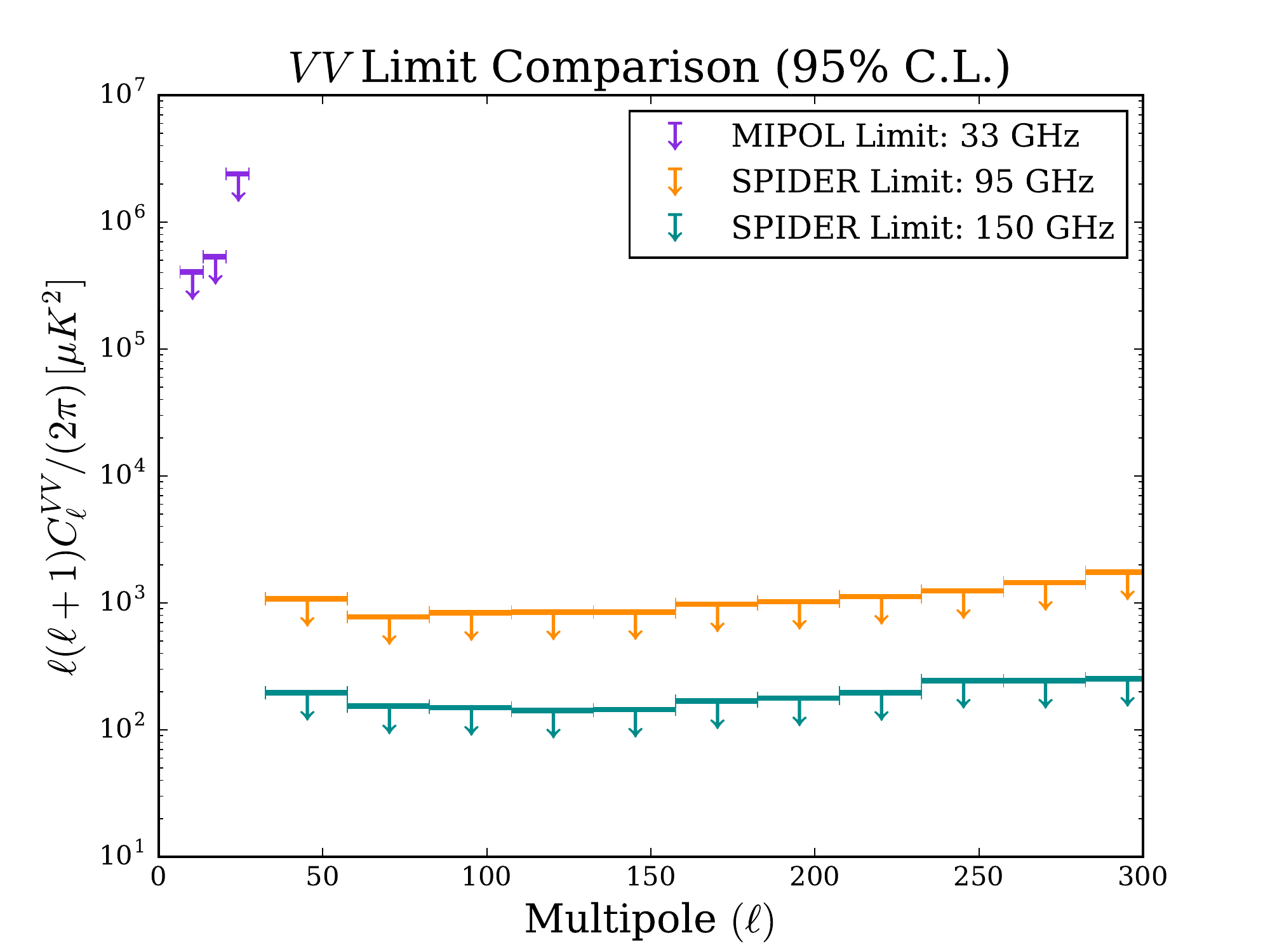}
\par
\end{centering}

\caption{\textsc{Spider}'s 95\% C.L. CMB circular polarization limits at 95 and 150 GHz. The MIPOL 33 GHz limit is also shown for comparison \citep{Mipol}.  \textsc{Spider}'s 150 GHz limit is stronger than the 95 GHz limit due to a combination of larger HWP circular polarization coupling and a larger number of detector channels and cross-spectra.  The numerical values of these limits are listed in Table \ref{table: limit_table}. Since the \textsc{Spider} limits assume a CMB source spectrum in the calculation of the $s$ parameters, these limits only apply to a thermal source.  When recalculated for $\nu^{-1}$ and $\nu^{-3}$ source spectra, the limits scale by 0.39 and 0.10 respectively at 95 GHz and 1.02 and 0.30 respectively at 150 GHz.  For synchrotron and thermal dust foreground models, these limits scale by 0.08 and 0.27 at 95 GHz and 0.11 and 0.56 at 150 GHz.}
\label{fig: 2sig_limits}
\end{figure*}

\begin{table}
\begin{center}
    \begin{tabular}{ *5l } 
    \toprule
    Bin Center ($\ell$) & 95 GHz Limit ($\mu K^2$) & 150 GHz Limit ($\mu K^2$) \\ \midrule
    45 & 1088 &	195 \\ 
    70 & 783  &	153 \\
    95 & 842  &	149 \\
    120 & 853  & 141 \\
    145 & 856  & 142 \\
    170 & 985  & 164 \\
    195 & 1032  & 177 \\ 
    220 & 1129  & 197 \\
    245 & 1254 & 242 \\
    270 & 1455 & 244 \\
    295 & 1760  & 255 \\
    \bottomrule
    \end{tabular}
    \caption{\textsc{Spider}'s 95\% C.L. limits on $\ell (\ell+1)C_{\ell}^{VV}/(2\pi)$ for a CMB source based on Figure \ref{fig: 2sig_limits}.}
    \label{table: limit_table}
\end{center}
\end{table}

Figure \ref{fig: V_spectra} shows \textsc{Spider}'s $VV$  CMB spectra at 95 and 150 GHz, neither of which indicate a significant detection of circular polarization.  The mean values and errors are derived from the distributions of the $s$-scaled cross-spectra, and the spread in each of those distributions has contributions from both the distribution of the various cross-spectra and the distributions of $s$ values.  Figure \ref{fig: 2sig_limits} shows the 95\% C.L. limits on CMB circular polarization derived from these spectra, and the numerical values are provided in Table \ref{table: limit_table}.  Although the measurements are made at different frequencies, they are expressed in units of CMB temperature, which are the equivalent fluctuations of a 2.73 K blackbody required to produce the measured intensity variations.  This result represents an improvement of several orders of magnitude over the previous best upper limit \citep{Mipol} at a complementary range of angular scales.  

However, \textsc{Spider}'s limits depend on the chosen source spectrum through the calculations of the HWP coupling parameters $s$.  Many of the methods for generating CMB circular polarization described in Section \ref{sec:intro} predict polarization signals with spectra of the form $\nu^{-1}$ or $\nu^{-3}$.  We therefore recalculate \textsc{Spider}'s $s$ distributions for such source spectra and find that the $VV$ limits in Figure \ref{fig: 2sig_limits} typically become lower.  Still expressed in CMB temperature units, they scale by factors of 0.39 and 0.10 respectively at 95 GHz and 1.02 and 0.30 respectively at 150 GHz.  In all of these cases, \textsc{Spider}'s circular polarization limits are still many orders of magnitude above the predicted cosmological signals. 

Similarly, the $VV$ limits presented in this paper can also be extended to upper limits on foreground circular polarization by recomputing $s$ with the appropriate source spectra.  \citet{Lubin_review} suggest that $\nu^{-3.4}$ is a reasonable model for synchrotron circular polarization.  With this source spectrum, \textsc{Spider}'s circular polarization limits in Figure \ref{fig: 2sig_limits} scale by 0.08 at 95 GHz and 0.11 at 150 GHz, still using CMB temperature units.  To obtain an estimate of \textsc{Spider}'s limit on the circular polarization of thermal dust, we use the linear polarization model of $\nu^{3.5}$ for the source spectrum \citep{planck_foregrounds} since we are not aware of any circularly polarized dust models.  This leads to circular polarization limits that scale from Figure \ref{fig: 2sig_limits} by 0.27 at 95 GHz and 0.56 at 150 GHz.  Although these models of the source spectra are relatively uncertain, the predicted $V$ foreground signals are many orders of magnitude below \textsc{Spider}'s sensitivity level.

\section{Conclusion}
\label{sec:discussion}
This paper presents a new upper limit on CMB circular polarization from $33<\ell<307$ at 95 and 150 GHz.  It was obtained by exploiting a non-ideality of the HWP polarization modulators used by \textsc{Spider} to measure linear polarization during a 2015 Antarctic flight. This represents an improvement of several orders of magnitude over the previous limit, providing 95\% C.L. constraints on $\ell (\ell+1)C_{\ell}^{VV}/(2\pi)$ ranging from 141 $\mu K ^2$ to 255 $\mu K ^2$ at 150 GHz for a thermal CMB spectrum.   When recalculated for $\nu^{-1}$ and $\nu^{-3}$ source spectra, this limit scales by 1.02 and 0.30 respectively.  Data from \textsc{Spider}'s second flight, planned for December 2018, could provide increased sensitivity at 95 and 150 GHz as well as a new measurement at 280 GHz over the same range of angular scales.

As linear polarization experiments become increasingly sensitive, the techniques described in this paper can be applied to provide stronger constraints on CMB circular polarization. Several current and planned experiments use either HWPs or Variable-delay Polarization Modulators (VPMs) \citep{VPM_paper}, both of which can be used to measure $V$.  Although the current limit is many orders of magnitude larger than the most optimistic signal predictions, these measurements provide an observational test of the standard cosmological model and a wide range of physical processes. Since this limit is still about four orders of magnitude above modern linear polarization measurements, a dedicated experiment with better $V$-coupling could make significant improvements using existing technology.

\acknowledgments
\textsc{Spider} is supported in the U.S. by the National Aeronautics and Space Administration under grants NNX07AL64G, NNX12AE95G, and NNX17AC55G issued through the Science Mission Directorate and by the National Science Foundation through PLR-1043515. Logistical support for the Antarctic deployment and operations was provided by the NSF through the U.S. Antarctic Program.  Support in Canada is provided by the Natural Sciences and Engineering Research Council and the Canadian Space Agency.  Support in Norway is provided by the Research Council of Norway.  Support in Sweden is provided by the Swedish Research Council through the Oskar Klein Centre (Contract No. 638-2013-8993).  K.F. acknowledges support from DoE grant DE-SC0007859 at the University of Michigan.  We also wish to acknowledge the generous support of the David and Lucile Packard Foundation, which has been crucial to the success of the project. 

The collaboration is grateful to the British Antarctic Survey, particularly Sam Burrell, for invaluable assistance with data and payload recovery after the 2015 flight.  We thank Brendan Crill and Tom Montroy for significant contributions to \textsc{Spider}'s development.  JMN wishes to thank Glenn Starkman for useful discussions about methods of generating CMB circular polarization. The computations described in this paper were performed on the GPC supercomputer at the SciNet HPC Consortium \citep{Scinet}. SciNet is funded by the Canada Foundation for Innovation under the auspices of Compute Canada, the Government of Ontario, Ontario Research Fund - Research Excellence, and the University of Toronto.

\bibliographystyle{apj}
%\bibliography{references}

\end{document}